# Calculation of The Critical Temperature for Anisotropic Two-Layer Ising Model Using The Transfer Matrix Method

*M. Ghaemi*[1]; *M. Ghannadi*[2] and B. Mirza[3]

Atomic Energy Organization of Iran, Deputy in Nuclear Fuel Production,
End of Karegar Av. Tehran, IRAN

[1] Corresponding author, Email: ghaemi@saba.tmu.ac.ir
[2] Email: mghannadi@seai.neda.net.ir
[3] Department of Physics, Isfahan University of Technology, Isfahan 84154, Iran
Email: b.mirza@cc.iut.ac.ir

ABSTRACT

A new finite-size scaling approach based on the transfer matrix method is developed to calculate the critical temperature of anisotropic two-layer Ising ferromagnet, on strips of $r$ wide sites of square lattices. The reduced internal energy per site has been accurately calculated for the ferromagnetic case, with the nearest neighbor couplings $K_x$, $K_y$ (where $K_x$ and $K_y$ are the nearest neighbor interactions within each layer in the $x$ and $y$ directions, respectively) and with inter-layer coupling $K_z$, using different size-limited lattices. The calculated energies for different lattice sizes intersect at various points when plotted versus the reduced temperature. It is found that the location of the intersection point versus the lattice size can be fitted on a power series in terms of the lattice sizes. The power series is used to obtain the critical temperature of the unlimited two-layer lattice. The results obtained, are in good agreement with the accurate values reported by others.

**Introduction**

The physical properties of various magnetic layered structures have been intensely studied both experimentally and theoretically for reasons ranging from fundamental investigations of phase transitions to technical problems encountered in thin-film magnets[1]. Although at zero magnetic field, there is an exact solution for the 2-dimensional Ising model[2], no such solution exists for the two-layer Ising model. The two-layer Ising model, as a simple generalization of the 2-D Ising model and also as an intermediate model between the 2-D and 3-D Ising models, has of long[3-6] been studied. Several approximation methods have been applied to this model [7-12].

Finite-size scaling (FSS) has bcome increasingly important in the study of critical phenomena[13, 14]. This is partly due to the progress in theoretical understanding of finite – size effects, and partly due to the application of FSS in the analysis of results from simulation methods. FSS allows us to extract some properties of the infinite system near a phase transition by studying finite, numerically accessible samples. Nightingale[15] has shown how the transfer matrix methods[16,17] could be made more powerful by combining them with FSS to extrapolate their results for finite systems to the thermodynamic limit. This technique has been applied with great success to the eight-vertex model[18], the square Ising antiferromagnet[19], the hard-square lattice gas[20], and the triangular Ising antiferromagnet[21]. In this method it is essential to compute the largest and the next largest eigenvalues of transfer matrix.

In a recent work, Ghaemi et al. [22] used the transfer matrix method in the numerical calculation of the critical temperature of Ising and three-state Potts ferromagnet models, on strips of $r$ wide sites of square, honeycomb and triangular lattices in the absence of a magnetic field. They have shown that the existence of the duality relation for the two-dimensional Ising model implies that the reduced internal energy per site, $u$ $(K)$, at the critical temperature must be independent of the size of lattice where $K=j/kT$ and $j>0$ is the coupling energy for the nearest neighbor spins. Owing to this fact, the calculated $u(K)$ curves plotted versus the reduced temperature, $K$, for the lattice with varying sizes, they were intersected at a single point known as the critical temperature. These authors were able to calculate the critical temperature of the square, triangular, and honeycomb lattices. They then extended the method to the 3-state Potts models to obtain the critical point for the 2-D lattices [22]. In yet another extension, they applied it to the 3-D Ising model [22]. Unlike the 2-D models, when $u(K)$ was plotted versus $K$ for the 3-D lattice, it was seen that the location of the intersection point depends on the lattice sizes, in other words there is no single intersection point. It was observed that the location of the intersection point versus the lattice sizes was almost linear. They extrapolated the line to infinite lattice size to obtain the critical temperature of a simple cubic lattice. The advantage of this method is that it is only sufficient to know the largest eigenvalue in order to compute the critical temperature.

In the present work, we have extended the method given in Ref. 22 to calculate the critical temperature numerically for the two-layer Ising model. In the following section, we will show that this method can be easily employed to obtain the critical point for the anisotropic two-layer Ising models. Similar to the 3-D Ising models, according to our results, the location of the intersection point depends on the lattice sizes, in other words there is no single intersection point. We have found that the location of the intersection point versus the lattice sizes can be fitted on a power series in terms of the lattice sizes. We have extrapolated the line to an infinite lattice size in order to obtain the critical temperature of the two-layer lattice.

**Calculation**

Consider a two-layer square lattice with the periodic boundary condition composed of slices, each with two layers, each layer with $p$ rows, where each row has $r$ sites. Each slice has then $2 \times p \times r = N$ sites and the coordination number of all sites is the same (namely 5). In the two-layer Ising model, for any site we define a spin variable $\sigma^{1(2)}(i, j) = \pm 1$, in such a way that $i=1,..., r$ and $j=1,..., p$, where subscript 1(2) denote the layer number. We include the periodic boundary condition as:

$$\sigma^{1(2)}(i + r, j) = \sigma^{1(2)}(i, j) \qquad (1)$$

$$\sigma^{1(2)}(i, j + p) = \sigma^{1(2)}(i, j) \qquad (2)$$

In this paper, we discuss the anisotropic ferromagnetic case with the nearest neighbor couplings $K_x$ and $K_y$, where $K_x$ and $K_y$ are the nearest neighbor interactions within each layer in the $x$ and $y$ directions, respectively, and with inter-layer coupling $K_z$.

We take only the interactions among the nearest neighbors into account. The configurational energy for the model may be defined as,

$$\frac{E(\sigma)}{kT} = -\sum_{i=1}^{r,*}\sum_{j=1}^{p,*}\sum_{n=1}^{2}\{K_x\sigma^n(i,j)\sigma^n(i+1,j) + K_y\sigma^n(i,j)\sigma^n(i,j+1)\}$$

$$- K_z\sum_{i=1}^{r}\sum_{j=1}^{p}\sigma^1(i,j)\sigma^2(i,j) \qquad (3)$$

where * indicates the periodic boundary conditions (Eqs.1, 2). The canonical partition function, Z (K), is

$$Z(K) = \sum_{\{\sigma\}} e^{\frac{-E(\sigma)}{kT}} \qquad (4)$$

Substitution of Eq.3 into Eq.4 gives,

$$Z(K) = \sum_{\sigma(\{i\},1)}\cdots\sum_{\sigma(\{i\},p)}\langle 1|B|2\rangle\langle 2|B|3\rangle\ldots\langle p|B|1\rangle \qquad (5)$$

where

$$|j\rangle = |\sigma^1(1,j)\rangle \otimes |\sigma^2(1,j)\rangle \otimes |\sigma^1(2,j)\rangle \otimes |\sigma^2(2,j)\rangle \ldots \otimes |\sigma^2(r,j)\rangle, \qquad (6)$$

$$\sum_{\sigma(\{i\},j)} = \sum_{\sigma^1(1,j)}\sum_{\sigma^1(2,j)}\cdots\sum_{\sigma^1(r,j)}\sum_{\sigma^2(1,j)}\sum_{\sigma^2(2,j)}\cdots\sum_{\sigma^2(r,j)}, \qquad (7)$$

and the element $B_{j,j+1}$ of the transfer matrix **B** is defined as,

$$B_{j,j+1} = \langle j|B|j+1\rangle = \exp[\sum_{i=1}^{r,*}\sum_{n=1}^{2}\{K_x\sigma^n(i,j)\sigma^n(i+1,j) + \rho K_x\sigma^n(i,j)\sigma^n(i,j+1)\}$$

$$+ \xi K_x\sum_{i=1}^{r}\sigma^1(i,j)\sigma^2(i,j)] \qquad (8)$$

where

$$\rho = \frac{K_y}{K_x} \qquad (9)$$

$$\xi = \frac{K_z}{K_x} \qquad (10)$$

By orthogonal transformation, the **B** matrix can be diagonalized, where Eq.4 for large values of $p$ can be written as [23]

$$Z(K) = (\lambda_{max})^p \tag{11}$$

where the $\lambda_{max}$ is the largest eigenvalue of **B**. From the well known thermodynamic relation for the Helmholtz free energy, $A = -kT\ln Z$, along with Eq.11, the following results are obtained:

$$a(K) = -\frac{A}{NkT} = \frac{\ln \lambda_{max}}{r} \tag{12}$$

$$u(K) = \frac{-E}{Nj} = \frac{\partial a(K)}{\partial K} \tag{13}$$

where $u(K)$ and $a(K)$ are the reduced internal energy and Helmholtz free energy per site, respectively. The numerical calculation of $u(K)$ can be easily programmed using the well-known mathematical software such as Maple, Mathlab, Mathematica,…, and ARPACK.

For the two-layer square lattice with size $r$, using Eq.8, the elements of the **B** matrix (of order $2^{2r}$) have been calculated numerically. The problem now reduces to calculating the largest eigenvalue of the **B** matrix for a semi-infinite strip of width $r$. Such a calculation becomes difficult when $r$ has large values (namely, larger than 5). In these cases, the size of the transfer matrix is larger than $2^{10} \times 2^{10}$ and hence the computation of the largest eigenvalue ($\lambda_{max}$) is difficult. We have employed the method in Refs.17 and 21 to reduce the transfer matrix size. Rather than diagonalizing the $2^{2r} \times 2^{2r}$ transfer matrix for $r = 3, 4, 5, 6, 7$, we diagonalize matrices of the rank 8, 22, 44, 135, 362, respectively. As seen, the size of the reduced transfer matrix is much smaller than that of the original transfer matrix **B**, such that the $\lambda_{max}$ can be easily calculated from the reduced matrix. The ARPACK package is used to diagonalize the reduced matrix, from which the $\lambda_{max}$ is calculated with a high precision for different values of $K_x$, $K_y$, and $K_z$. The calculated reduced internal energy per site for the two-layer square lattice with ($r = 3, 4,…, 7$) was plotted versus $K_x$, for the case in which $\xi = K_z / K_x = 1.3$ and $\rho = K_y / K_x = 0.4$, as shown in Figure 1. It can be seen in this figure, that the location of the intersection point depends on the size of the lattices ($r$ and $r'$). It is also evident from Figure 1 that the value of $K$ at the intersection point ($K_n$) decreases when the lattice size increases. It is necessary to determine the intersection point for two unlimited lattices with different sizes in order to obtain the value of the critical temperature $K_C$. However, such a point may be predicted if we have an expression for the intersection point in terms of $1/n$, where $n = 4rr'$. To do so, the intersection points shown in Figure 1 were plotted versus $1/n$. As shown in Figure 2, the points may be fitted into a power series in terms of $1/n$. For this purpose, we have used a polynomial of degree 3 and the coefficients have been calculated, using the least square method. The results obtained are as follows:

$$K_n = \sum_{j=0}^{3} a_j \left(\frac{1}{n}\right)^j$$

$K_n = 0.4173 + 0.708\,(1/n) - 20.6\,(1/n)^2 + 1.073\times10^{-7}\,(1/n)^3$

(14)

If Eq.12 is applicable for large lattice sizes, then for the limit $n \to \infty$, $K_c$ is equal to $a_0$. Such a calculation had also been performed for different values of $\xi$ and $\rho$. The results from these calculations are given in Table 1. As shown in Table1, the value of critical temperature is quite sensitive to the value of $\xi$ and $\rho$.

**Conclusion**

We have shown that our simple approach can be used to calculate the critical temperature of the anisotropic two-layer Ising model. As it is seen, the results obtained are in good agreement with the values obtained from the recent corner transfer matrix renormalization group (CTMRG) method[12]. The small inaccuracy in our results is due to the limitation dictated by available computer resource. However, we expect to obtain even higher accuracy levels, if matrices of higher orders than $2^{16}$ could be used. As the simple approach introduced in this paper was capable of being successfully applied to the two-layer Ising model, 2-D Ising and Potts models, and 3-D Ising model [22], we may expect that it can also be used in the case of other lattice models.

**TABLE CAPTION**
Table1
The calculated critical temperatures of the two-layer Ising model compared to the other results, for different values of $\xi = \dfrac{K_z}{K_x}$, $\rho = \dfrac{K_y}{K_x}$.

| $\xi$ | $\rho$ | | | | | |
|---|---|---|---|---|---|---|
| | 0.1 | 0.4 | 0.7 | 1.0 | | 1.3 |
| 0.1 | 0.879 | 0.582 | 0.464 | 0.397 | 0.3977 [1] | 0.348 |
| 0.4 | 0.763 | 0.510 | 0.713 | 0.354 | 0.3541 [1] | 0.315 |
| 0.7 | 0.686 | 0.465 | 0.381 | 0.330 | * | 0.293 |
| 1.0 | 0.651 | 0.436 | 0.359 | 0.311 | 0.3117 [1] | 0.277 |
| 1.3 | 0.629 | 0.417 | 0.343 | 0.298 | * | 0.267 |

(1) From the recent corner transfer matrix renormalization group (CTMRG) method given in Ref.12.

**FIGURE CAPTIONS**

Figure. 1

The reduced internal energy versus the reduced temperature, $K_x = J_x/kT$, for the two-layer Ising model with different lattice size for $\xi = 1.3$ and $\rho = 0.4$.

Figure. 2

The intersection points of Figure 1($\bullet$) versus $1/n$ which is fitted into a polynomial with degree of three (solid line).

Fig. 1

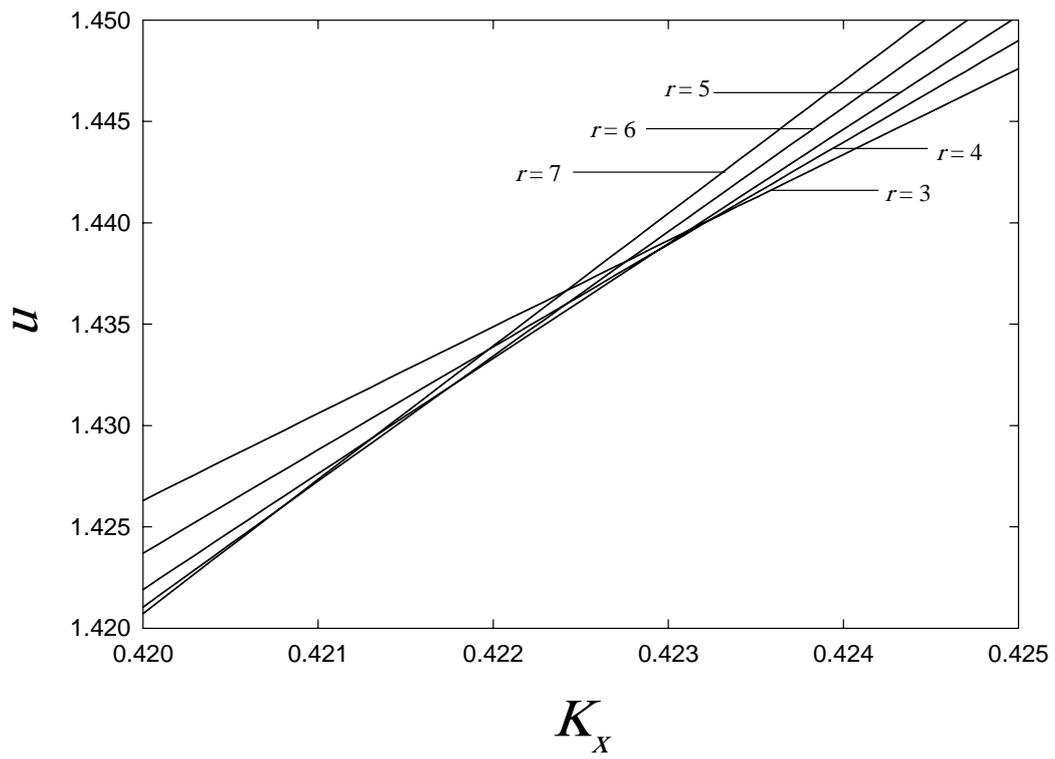

Fig. 2

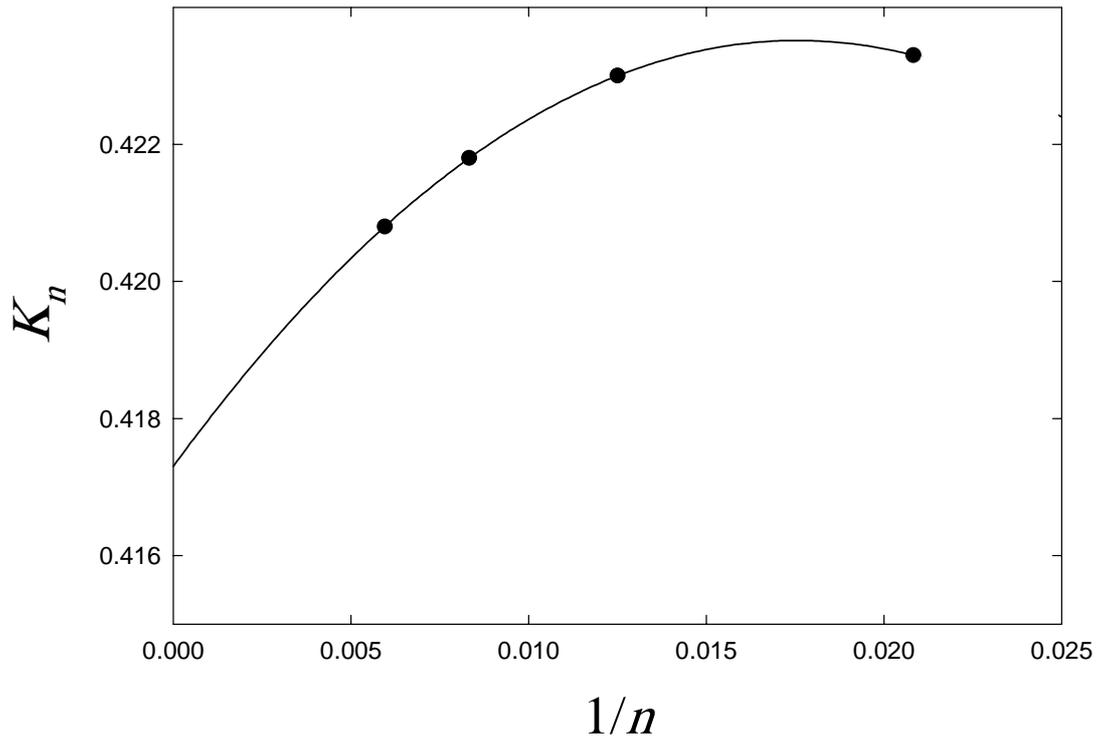